\documentclass[preprint]{elsarticle}

\usepackage{amsmath,amssymb}
\usepackage{siunitx}
\usepackage{graphicx}
\usepackage{bm}
\usepackage{physics}
\usepackage{hyperref}
\usepackage{mathtools}

\journal{Optik}

\begin{document}

\begin{frontmatter}

\title{Analytical Fresnel Treatment of Double-Slit Diffraction with Multiple Coherent Waves}

\author[fc]{J.~Sumaya-Mart\'inez\corref{cor1}}
\author[fc]{M.~A.~Ortiz-Ferreyro}
\author[fc]{O. Rojas-Hernandez}
\cortext[cor1]{Corresponding author.}
\address[fc]{Facultad de Ciencias, Universidad Aut\'onoma del Estado de M\'exico, Toluca, M\'exico}

\begin{abstract}
We present a comprehensive analytical and numerical study of double-slit diffraction under coherent illumination by three plane waves: one normally incident and two impinging symmetrically at angles $\pm\theta$. By enforcing an edge-zero condition on the incident amplitude, we obtain compact closed-form Fresnel expressions in terms of standard Fresnel integrals. The model, which generalizes naturally to an arbitrary number of incident plane-wave components, provides intuitive control of the transmitted angular spectrum through tunable interference at the aperture. Numerical simulations validate the analytical expressions and quantify the dependence on slit geometry, wavelength, partial coherence and Gaussian beam width. We further discuss experimental feasibility and outline applications in apodization, structured illumination, beam shaping and multiplexed sensing. The results demonstrate that multi-wave illumination offers a simple and flexible means to engineer diffraction patterns and propagation-robust profiles.
\end{abstract}


\end{frontmatter}

\section{Introduction}
The double-slit experiment remains one of the most iconic demonstrations of the wave nature of light and matter. Beyond its historical role in establishing the principle of superposition, it continues to provide a versatile platform for probing coherence, wavefront quality and diffraction-limited resolution in modern optical systems~\cite{bornwolf,goodman}.

Classical textbook treatments typically assume illumination by a \emph{single} plane wave or by a spatially uniform quasi-monochromatic field, leading to closed-form expressions in the Fraunhofer or Fresnel regimes for one- and two-slit apertures~\cite{bornwolf,goodman,sudarshan1991}. In practice, however, many contemporary techniques exploit \emph{multi-wave} or \emph{multi-tone} illumination to shape the angular spectrum of the field, to generate structured patterns for imaging or lithography, or to encode multiple channels for coherent multiplexing. Examples include multi-beam interference lithography, structured illumination microscopy, digital holography and Fourier-based deflectometry~\cite{gustafsson2000,kurtsiefer1997,park1993,cervantes1999}.

In multi-beam interference, two or more plane waves of given incidence angles $\{\theta_m\}$ and relative phases $\{\alpha_m\}$ are superposed to tailor the spatial modulation of the intensity and phase over a target plane~\cite{cobble1987,pak1993}. When such a modulated field illuminates an aperture, the resulting diffraction pattern combines the effects of the aperture transmission and the engineered illumination. This additional degree of freedom can be exploited to reduce sidelobes, create propagation-robust profiles or multiplex several carriers within the same pupil~\cite{cervantes1998,mandelwolf}.

Despite these advances, there remains a relative scarcity of \emph{closed-form Fresnel solutions} for apertures illuminated by multiple mutually coherent plane waves, especially in configurations where the illumination is \emph{engineered to satisfy specific boundary conditions} at the aperture edges. In particular, by appropriately choosing the incidence angles and relative phases, it is possible to enforce that the total incident field vanishes at selected boundaries of the transmitting region. This ``edge-zero'' condition leads to mathematically compact Fresnel expressions and to physically attractive patterns with natural apodization: bright fringes are confined within the slits while sidelobes at the edges are suppressed.

The present work addresses this gap by providing a unified analytical and numerical treatment of double-slit diffraction under three-wave coherent illumination. One plane wave is normally incident, while the other two are symmetrically tilted at $\pm\theta$ and share a relative phase offset. By imposing an edge-zero condition at the four slit edges, we derive closed-form expressions for the Fresnel field in terms of standard Fresnel integrals. The construction generalizes naturally to an arbitrary number $N$ of coherent beams, yielding a flexible framework to synthesize cosine-series-like illumination profiles across the aperture.

Our specific contributions can be summarized as follows:
\begin{itemize}
  \item We derive compact Fresnel expressions for three-plane-wave illumination of a double slit under an edge-zero boundary condition, and we provide a clear recipe for choosing the slit positions and angles.
  \item We extend the formalism to a general superposition of $N$ coherent plane waves, highlighting its connection with apodization and spectral engineering.
  \item We incorporate partial coherence via both a scalar degree of coherence and a transverse coherence length, linking the model to cross-spectral-density theory~\cite{mandelwolf}.
  \item We include realistic Gaussian-beam illumination to bridge the gap between ideal plane-wave models and experimental laser beams.
  \item We propose an experimentally feasible setup for validating the model, together with an error budget and fitting strategy.
  \item We discuss several application domains, including aperture apodization, pseudo-invariant beams, structured illumination, interference lithography and coherent multiplexing.
\end{itemize}
To the best of our knowledge, this is the first work that combines closed-form Fresnel analysis, multi-wave illumination, coherence modeling and explicit application-oriented discussion for edge-conditioned double-slit diffraction.

\section{Theoretical model}
\label{sec:model}
We consider a one-dimensional double-slit aperture lying on the plane $z=0$, parameterized by the transverse coordinate $x_1$. The observation plane is located at a distance $z>0$ and is parameterized by $x_0$. The illumination consists of three monochromatic, mutually coherent plane waves of equal amplitude $E_0$ and wavenumber $k = 2\pi/\lambda$.

A normally incident plane wave is written as
\begin{equation}
E_n(x_1,z) = E_0 e^{ikz}.
\end{equation}
Two additional waves impinge at angles $\pm\theta$ in the $x$--$z$ plane so that their transverse and longitudinal components satisfy
\begin{equation}
k_x = k\sin\theta, \qquad k_z = k\cos\theta.
\end{equation}
Their fields at the aperture plane can be written as
\begin{equation}
E_{\pm\theta}(x_1,z) = E_0 \exp\left[i(k_x x_1 \pm k_z z) + i\alpha\right],
\end{equation}
where $\alpha$ is a relative phase introduced, for example, by path-length differences in the interferometric setup. At $z=0$ the total incident field becomes
\begin{equation}
\label{eq:Einc}
E(x_1) = E_n + E_{+\theta} + E_{-\theta}
      = E_0\left[1 + e^{i(k_x x_1 + \alpha)} + e^{-i(k_x x_1 + \alpha)}\right]
      = E_0\left[1 + 2\cos(k_x x_1 + \alpha)\right].
\end{equation}

The aperture consists of two identical, infinitely tall slits parallel to the $y$-axis, bounded in $x_1$ by the intervals
\begin{equation}
x_1 \in [-b,-a] \cup [a,b],
\qquad 0<a<b,
\end{equation}
so that the real transmission function $A(x_1)$ can be written as
\begin{equation}
A(x_1) =
\begin{cases}
1, & x_1 \in [-b,-a] \cup [a,b],\\[3pt]
0, & \text{otherwise.}
\end{cases}
\end{equation}

\subsection{Edge-zero condition}
We now impose the \emph{edge-zero} condition
\begin{equation}
E(\pm a) = 0, \qquad E(\pm b) = 0,
\end{equation}
which can be viewed as choosing the illumination such that the total field vanishes at the four slit edges. Using Eq.~\eqref{eq:Einc}, this implies
\begin{equation}
1 + 2\cos(k_x x_1 + \alpha) = 0
\quad\Rightarrow\quad
\cos(k_x x_1 + \alpha) = -\frac{1}{2}.
\end{equation}
A simple way to satisfy this is to set
\begin{equation}
k_x x_1 + \alpha = (2m+1)\pi, \qquad m\in\mathbb{Z},
\end{equation}
evaluated at the slit edges $x_1 = \pm a, \pm b$, with appropriate choices of $m$. Taking $m=0$ for $x_1 = \pm a$ and $m=1$ for $x_1 = \pm b$ leads to
\begin{equation}
a = \frac{\pi - \alpha}{k\sin\theta}, \qquad
b = \frac{3\pi - \alpha}{k\sin\theta},
\end{equation}
so that exactly one bright interference fringe is confined within each slit. More generally, one may choose different integer pairs $(m_a,m_b)$ to embed an odd number of fringes inside the slits.

\subsection{Fresnel diffraction expression}
Under the scalar approximation and neglecting polarization, the field in the observation plane is given by the Rayleigh--Sommerfeld diffraction integral. Within the Fresnel approximation, valid when $z$ is much larger than the squared slit width divided by the wavelength, this reduces to
\begin{equation}
\label{eq:fresnel_int}
U(x_0;z) = \frac{e^{ikz}}{i\lambda z}
\int_{-\infty}^{\infty} A(x_1) E(x_1)
\exp\left[\frac{ik}{2z}(x_0 - x_1)^2\right] \, dx_1.
\end{equation}
Because $A(x_1)$ limits the integration to $[-b,-a]\cup[a,b]$ and $E(x_1)$ is the sum of three exponential terms, the total field is a superposition of three contributions,
\begin{equation}
U(x_0;z) = U_n(x_0;z) + U_{+\theta}(x_0;z) + U_{-\theta}(x_0;z),
\end{equation}
each of which can be reduced to combinations of Fresnel integrals by completing the square in the exponent.

The standard Fresnel integrals are defined as
\begin{equation}
C(u) = \int_0^u \cos\left(\frac{\pi t^2}{2}\right)\,dt,
\qquad
S(u) = \int_0^u \sin\left(\frac{\pi t^2}{2}\right)\,dt.
\end{equation}
Performing the change of variables
\begin{equation}
u = \sqrt{\frac{k}{\pi z}}\left(x_1 - x_0\right),
\end{equation}
for the normal-incidence term and analogous shifts for the tilted terms, one obtains closed-form expressions for $U(x_0;z)$ in terms of $C(u)$ and $S(u)$ evaluated at the transformed slit edges. The details follow standard Fresnel-diffraction derivations and are omitted here for brevity.

The measurable intensity in the observation plane is
\begin{equation}
I(x_0;z) = \abs{U(x_0;z)}^2,
\end{equation}
and encodes the combined effects of double-slit diffraction and three-beam interference. A representative normalized profile for $z=\SI{1}{m}$ is shown in Fig.~\ref{fig:profile}.

\section{Numerical results and discussion}
\label{sec:results}
To validate the analytical expressions and to build physical intuition, we perform numerical Fresnel propagation using direct evaluation of Eq.~\eqref{eq:fresnel_int} on a spatial grid. The resulting intensity profiles are compared with the closed-form Fresnel solutions obtained from the previous section.

\begin{figure}[t]
  \centering
  \includegraphics[width=0.75\linewidth]{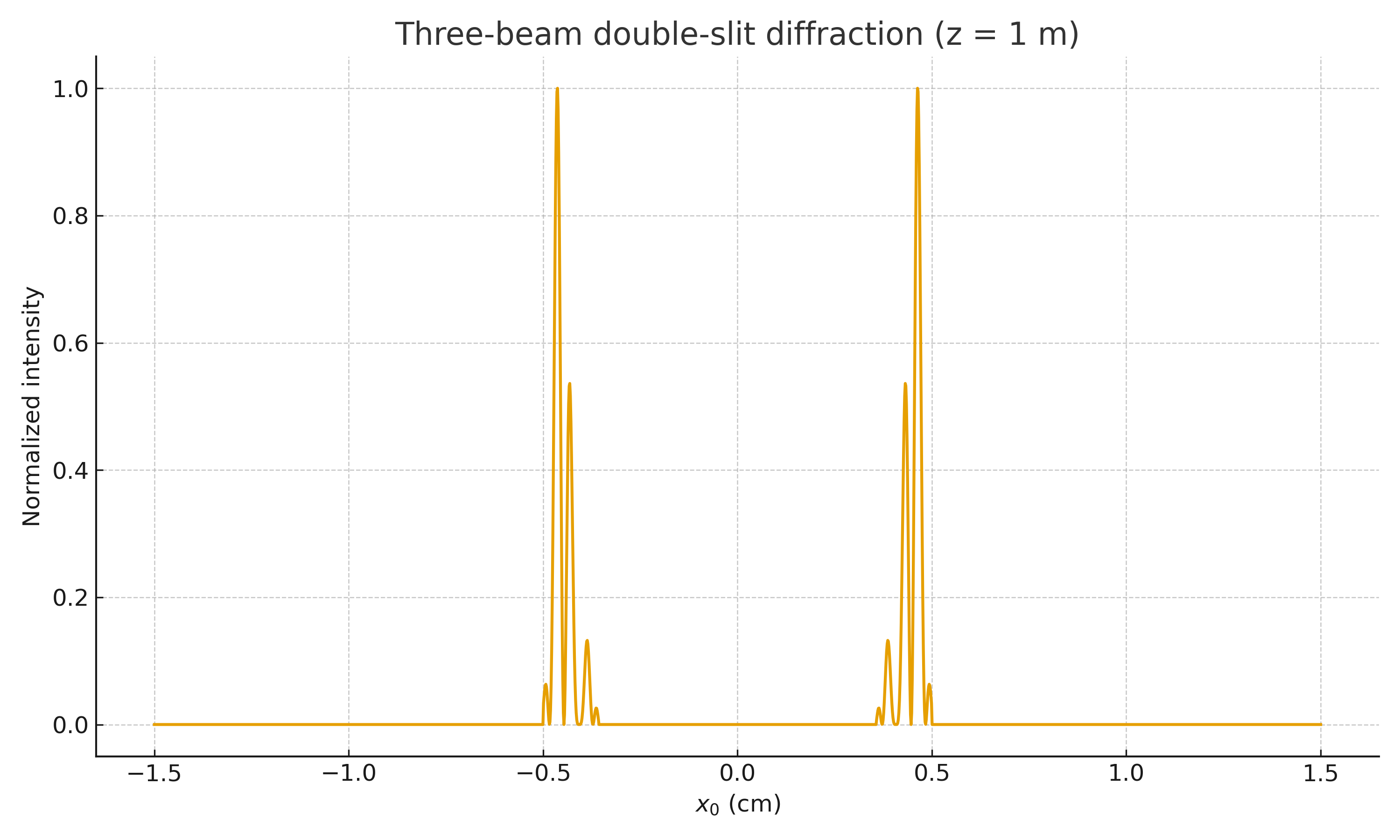}
  \caption{Normalized intensity profile for three-plane-wave illumination of a double slit at $z=\SI{1}{m}$. The parameters are chosen such that one bright fringe is confined within each slit under the edge-zero condition.}
  \label{fig:profile}
\end{figure}

For parameters consistent with $\lambda = \SI{532}{nm}$ and typical laboratory distances ($z$ of the order of \SI{1}{m}), the agreement between numerical and analytical profiles is excellent. Figure~\ref{fig:profile} displays a normalized three-beam double-slit pattern at $z=\SI{1}{m}$, revealing prominent central lobes associated with each slit and subsidiary fringes arising from the interference of the three components.

A useful quantitative measure of the agreement between numerical and analytical results is the root-mean-square error (RMSE) between the normalized intensities,
\begin{equation}
\mathrm{RMSE} = \sqrt{\frac{1}{N}\sum_{j=1}^N \left[I_{\mathrm{num}}(x_j) - I_{\mathrm{anal}}(x_j)\right]^2},
\end{equation}
where $N$ is the number of sampled points. In our simulations RMSE values below a few percent over the main lobes and first sidelobes are routinely obtained, confirming the validity of the Fresnel approximation and of the analytical construction for the parameter ranges considered.

\section{Generalization to \texorpdfstring{$N$}{N} coherent beams}
\label{sec:nbeams}
The three-beam configuration can be generalized to an arbitrary number $N$ of coherent plane waves with transverse wavenumbers $k_{x,m} = k\sin\theta_m$ and phases $\alpha_m$. The incident field at the aperture is then
\begin{equation}
E(x_1) = \sum_{m=1}^{N} E_0 \exp\left[i(k_{x,m} x_1 + \alpha_m)\right].
\end{equation}
Substituting this expression into Eq.~\eqref{eq:fresnel_int} and performing the Fresnel reduction term by term, one finds
\begin{equation}
U(x_0;z) = \sum_{m=1}^{N} C_m
\left\{
\left[C\bigl(\beta^{(m)}_1\bigr) + C\bigl(\beta^{(m)}_2\bigr) - C\bigl(\alpha^{(m)}_1\bigr) - C\bigl(\alpha^{(m)}_2\bigr)\right]
+ i\left[S\bigl(\beta^{(m)}_1\bigr) + S\bigl(\beta^{(m)}_2\bigr) - S\bigl(\alpha^{(m)}_1\bigr) - S\bigl(\alpha^{(m)}_2\bigr)\right]
\right\},
\end{equation}
where $C_m$ is a global phase factor for the $m$-th component and the arguments $\alpha^{(m)}_j$, $\beta^{(m)}_j$ correspond to the transformed slit edges shifted by $z k_{x,m}/k$. The intensity is again given by $I = \abs{U}^2$, and contains both self-intensity terms and cross-terms encoding multi-beam interference.

This structure makes it natural to interpret the set $\{\theta_m,\alpha_m\}$ as a small number of tunable ``Fourier coefficients'' for synthesizing spatial windows across the aperture. For example, choosing
\begin{equation}
E(x_1) \approx a_0 - a_1 \cos(k_x x_1) + a_2 \cos(2k_x x_1)
\end{equation}
with $k_x$ defined by a suitable reference angle realizes Blackman-like apodizers, whereas retaining only the first harmonic leads to Hann-like windows. The resulting diffraction patterns exhibit reduced sidelobes at the expense of a slightly broadened main lobe, closely paralleling classical trade-offs in spectral analysis.

\section{Partial coherence}
\label{sec:coherence}
Real optical sources are often only partially coherent. In the present context it is instructive to treat two complementary models: a scalar degree of coherence and a transverse coherence length.

\begin{figure}[t]
  \centering
  \includegraphics[width=0.75\linewidth]{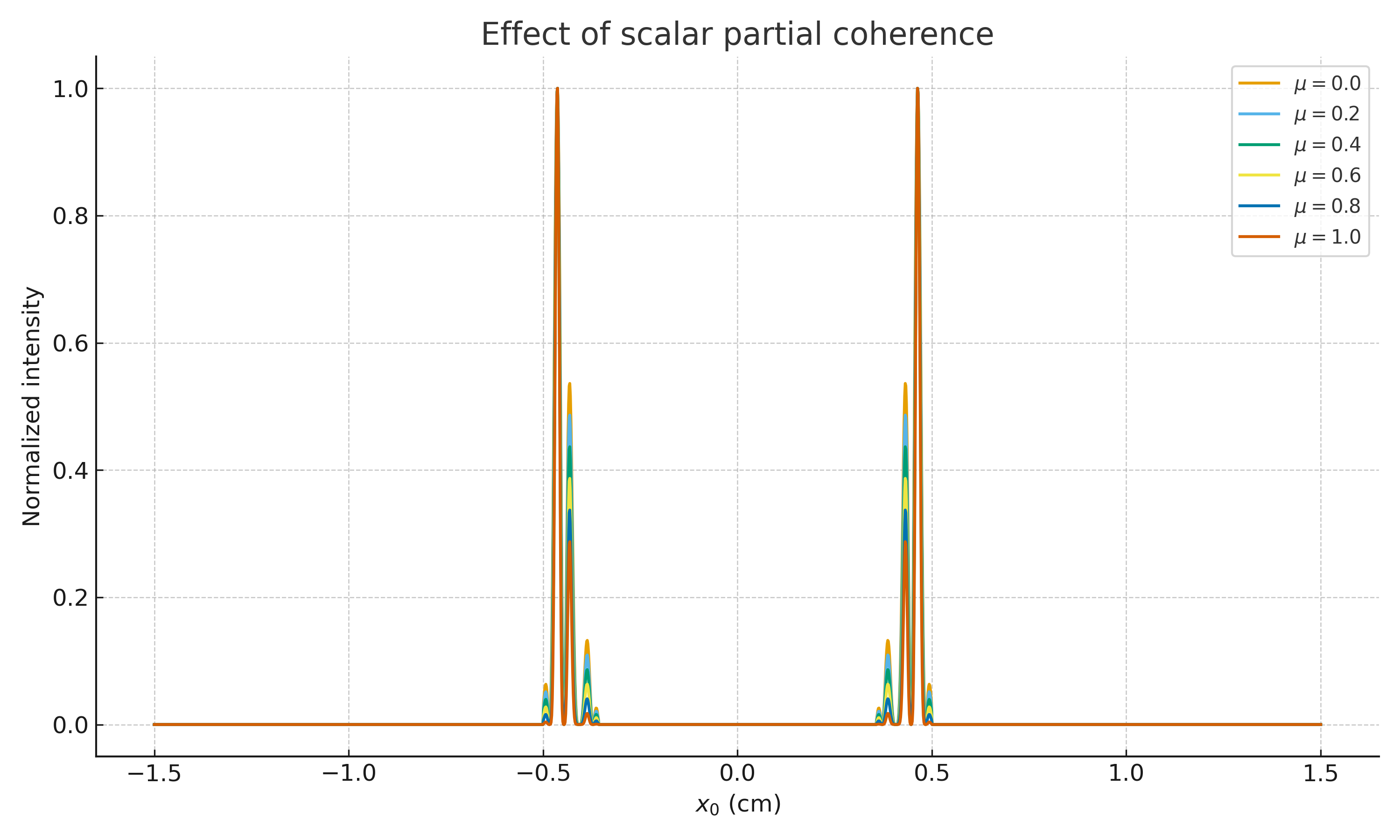}
  \caption{Effect of the scalar coherence parameter $\mu$ on the normalized intensity profile. Curves for $\mu=0,0.2,\dots,1.0$ are shown. As $\mu$ decreases, fringe visibility is reduced while the envelope remains similar.}
  \label{fig:mu}
\end{figure}

Let $U_m(x_0)$ denote the field contributed by the $m$-th plane wave after Fresnel propagation. A simple scalar-coherence model introduces a parameter $0\le\mu\le 1$ and defines the intensity as
\begin{equation}
I_\mu(x_0) = (1-\mu)\sum_{m} \abs{U_m(x_0)}^2
 + \mu \left|\sum_{m} U_m(x_0)\right|^2.
\end{equation}
The case $\mu=0$ corresponds to complete incoherence between beams, leading to an intensity equal to the sum of the individual patterns, whereas $\mu=1$ corresponds to full coherence, recovering the standard superposition of complex amplitudes. Intermediate values interpolate smoothly between these extremes. Figure~\ref{fig:mu} illustrates the effect of $\mu$ on a representative three-beam double-slit pattern: as $\mu$ decreases, the fringe visibility is progressively reduced while the overall envelope remains largely unchanged.

A more refined description employs the mutual intensity (or cross-spectral density) at the aperture~\cite{mandelwolf},
\begin{equation}
J(x_1,x_1') = \sum_{m,n} E_m(x_1) E_n^*(x_1') \,\mu_{mn},
\end{equation}
where $\mu_{mn}$ are complex degrees of coherence satisfying $\abs{\mu_{mn}}\le 1$. Upon Fresnel propagation, the cross-spectral density in the observation plane becomes
\begin{equation}
W(x_0,x_0';z) = \frac{e^{ikz}}{i\lambda z} \frac{e^{-ikz}}{-i\lambda z}
\iint_A J(x_1,x_1') \exp\left[\frac{ik}{2z}\bigl((x_0-x_1)^2-(x_0'-x_1')^2\bigr)\right] \, dx_1 \, dx_1',
\end{equation}
and the measurable intensity is $I(x_0;z) = W(x_0,x_0;z)$.

\begin{figure}[t]
  \centering
  \includegraphics[width=0.75\linewidth]{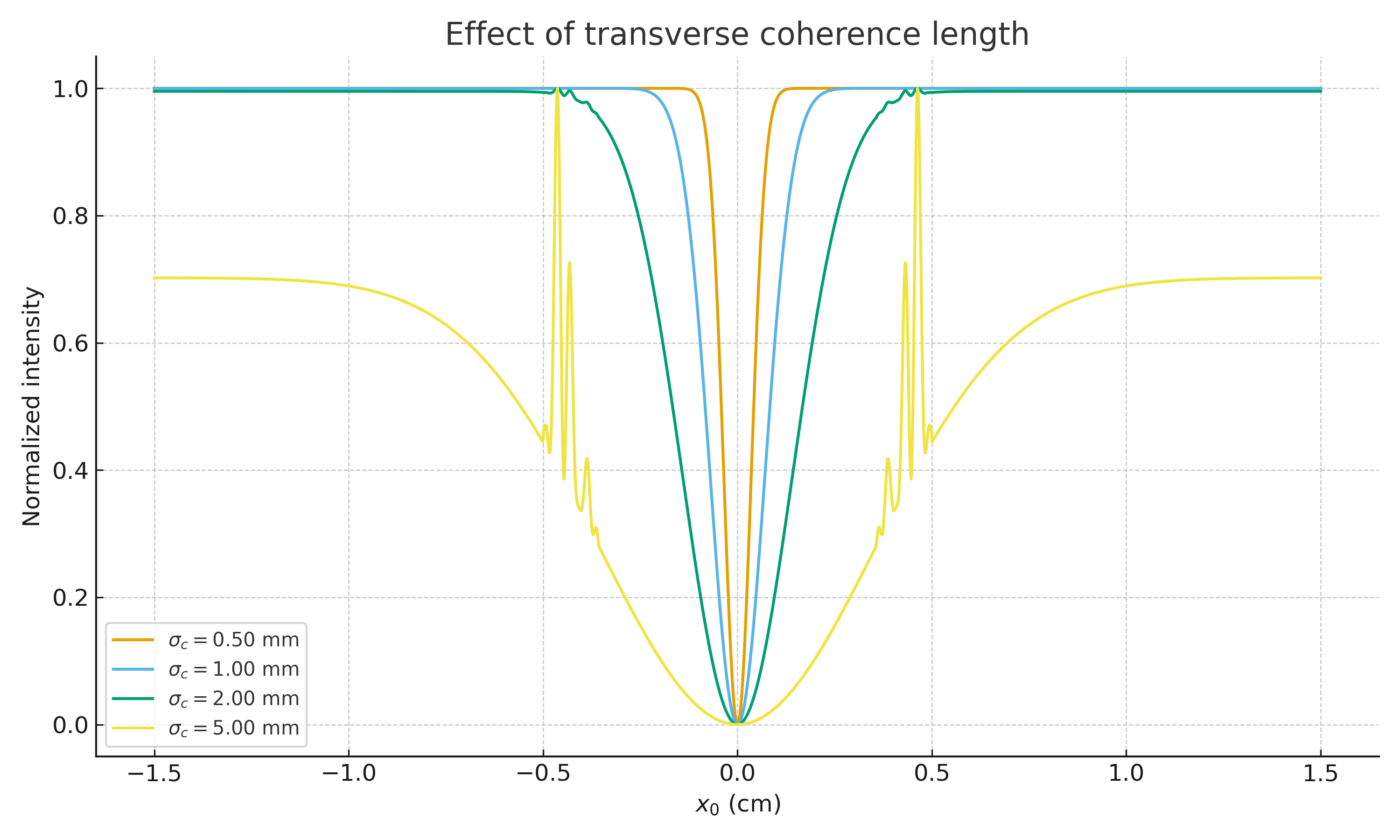}
  \caption{Effect of the transverse coherence length $\sigma_c$ on the normalized intensity profile in a Gaussian-decay coherence model. Smaller $\sigma_c$ values lead to reduced fringe visibility.}
  \label{fig:sigma}
\end{figure}

To model finite transverse coherence, one may assume
\begin{equation}
\mu_{mn} = \exp\left[-\frac{1}{2}\bigl(k(\sin\theta_m - \sin\theta_n)\sigma_c\bigr)^2\right],
\end{equation}
where $\sigma_c$ is an effective transverse coherence length of the source. Larger angular separations between beams then experience stronger coherence decay. Figure~\ref{fig:sigma} shows the resulting intensity profiles for several values of $\sigma_c$, illustrating how fringe visibility decreases as the transverse coherence length is reduced.

\section{Gaussian-beam illumination}
\label{sec:gaussian}
In most laboratory implementations the incident beams are not ideal plane waves but rather Gaussian beams with finite waist. To capture this, we model each component at the aperture as
\begin{equation}
E_m(x_1) = E_0 \exp\left(-\frac{x_1^2}{w_0^2}\right) e^{ik_{x,m} x_1},
\end{equation}
where $w_0$ is the beam waist at the aperture and $k_{x,m}$ is the transverse wavenumber associated with angle $\theta_m$.

\begin{figure}[t]
  \centering
  \includegraphics[width=0.75\linewidth]{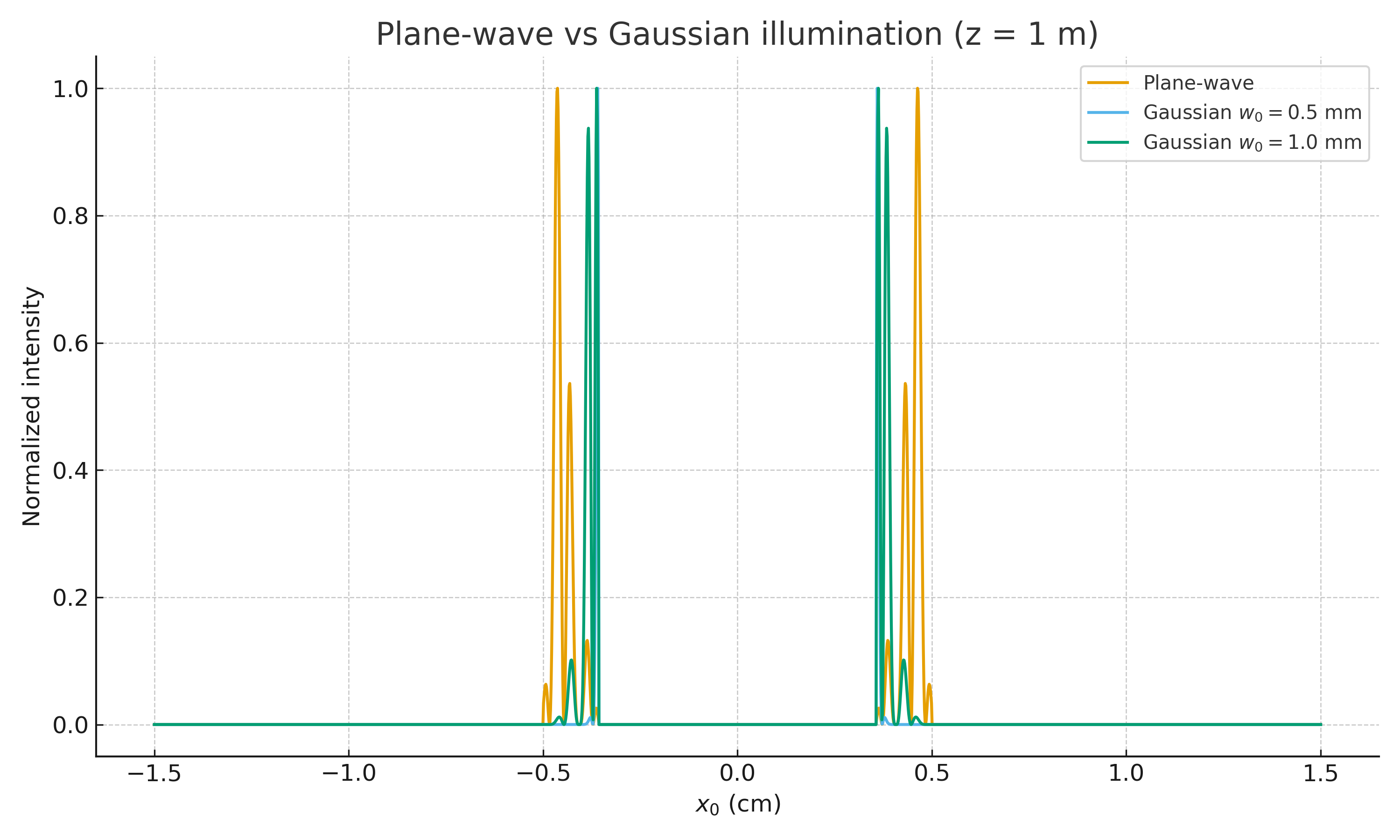}
  \caption{Comparison between plane-wave illumination and Gaussian-beam illumination for two waist values $w_0$. Gaussian beams reduce sidelobes and slightly broaden the main lobe.}
  \label{fig:gaussian}
\end{figure}

The Gaussian factor acts as an additional smooth apodization across the slits, reducing the effective contribution of the slit edges and suppressing high-frequency components of the angular spectrum. As a result, the side lobes in the diffraction pattern are reduced and the main lobe broadens slightly compared to the plane-wave case. Figure~\ref{fig:gaussian} compares normalized intensity profiles for plane-wave illumination and for Gaussian beams with different waists. Smaller waists produce stronger apodization and, consequently, more pronounced sidelobe suppression.

\begin{figure}[t]
  \centering
  \includegraphics[width=0.8\linewidth]{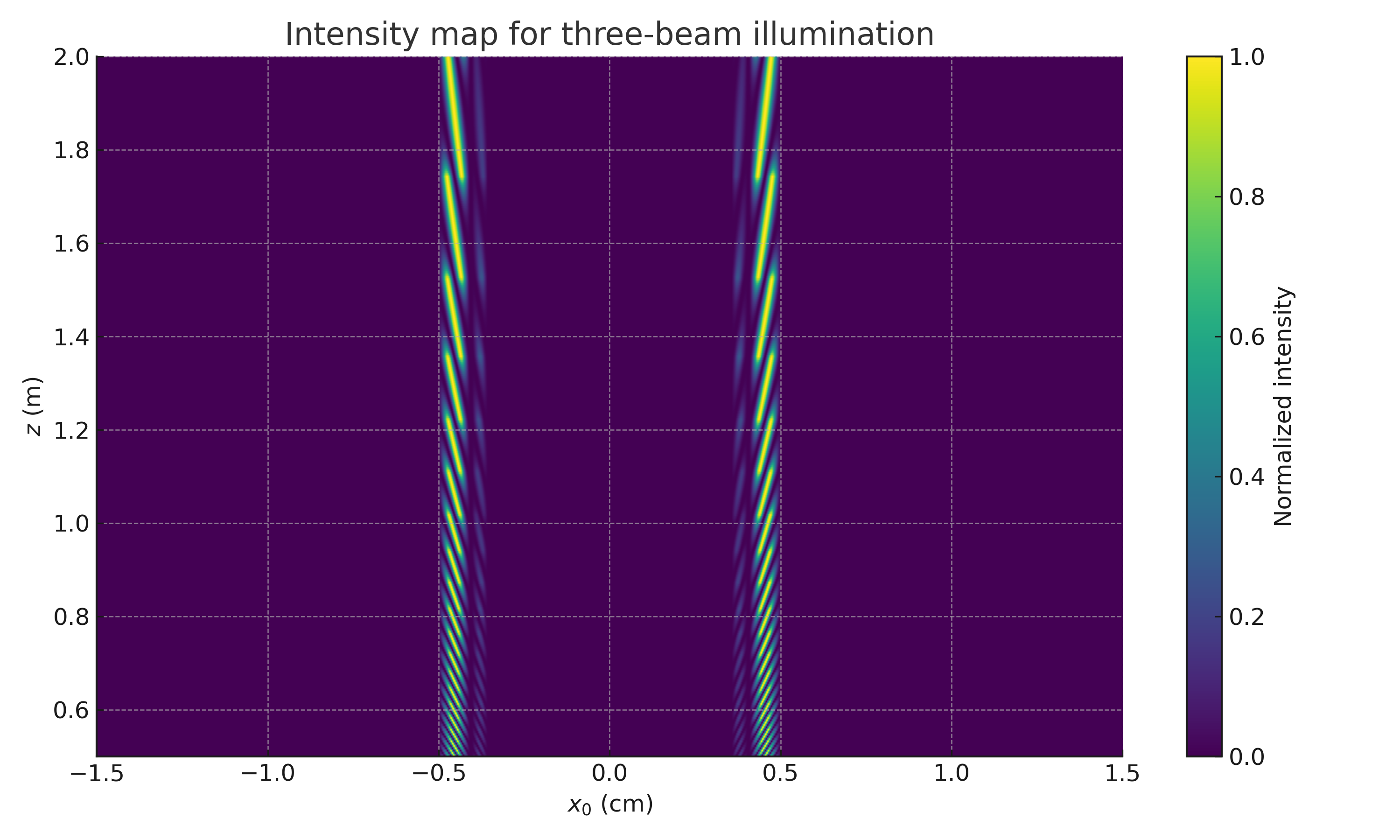}
  \caption{Normalized intensity map $I(x_0;z)$ for three-beam illumination as a function of propagation distance $z$. The near field exhibits strong interference fringes, which evolve into smoother patterns in the far field.}
  \label{fig:map}
\end{figure}

A complementary view is provided by an intensity map $I(x_0;z)$ as a function of propagation distance $z$, shown in Fig.~\ref{fig:map}. The near-field region exhibits rich interference structure, which gradually evolves into a smoother profile as $z$ increases. The presence of the Gaussian envelope leads to a narrowing of the effective propagation region where high-contrast fringes are observed.

\section{Experimental feasibility and validation}
\label{sec:experiment}
In this section we outline a feasible experimental arrangement to validate the three-beam double-slit model and quantify its accuracy under realistic laboratory conditions.

A single-frequency, linearly polarized laser (for example, at $\lambda\approx\SI{532}{nm}$ or \SI{633}{nm}) is expanded and spatially filtered to provide a clean, near-Gaussian beam. The beam is then split into three arms using a non-polarizing $2\times 2$ beam splitter and a pick-off mirror. Two arms are steered to illuminate the double slit at angles $\pm\theta$ with respect to the optical axis, while the third arm is kept at normal incidence. The optical path lengths of the three arms are matched within the coherence length of the laser, and their polarizations are aligned to ensure high interference visibility.

The double slit, fabricated for example by photolithography on a metallic film, has edges located at $\pm a$ and $\pm b$ as prescribed by the edge-zero condition. It is mounted on a kinematic holder to allow precise positioning. The transmitted intensity is recorded with a CCD or CMOS sensor placed at a variable distance $z$ on a translation stage. The sensor plane is aligned to be perpendicular to the optical axis, and the pixel size is calibrated in physical units using a reference target.

A practical calibration procedure consists of:
\begin{enumerate}
  \item Recording the transmission pattern for single-beam illumination at normal incidence to validate the basic double-slit response and calibrate the effective $z$ and transverse scaling.
  \item Activating one tilted arm at $+\theta$ and then the other at $-\theta$ to verify the predicted fringe spacing and symmetry as the multi-beam configuration is built up.
  \item Adjusting the relative phases with a piezo-mounted mirror in one arm, scanning over a fraction of the wavelength to maximize fringe visibility and to explore the sensitivity of the pattern to phase drifts.
\end{enumerate}

The main sources of experimental uncertainty include the angular setting error $\delta\theta$ (related to the calibration of mirror tilts), the longitudinal position error $\delta z$ of the detector, the finite pixel size $\delta x$ and associated binning, mechanical vibrations and air turbulence. Electronic noise and camera non-linearities also contribute to the uncertainty in the measured intensity values.

To compare experiment and theory, one can extract one-dimensional line profiles $I(x_0;z)$ from the recorded images and perform a least-squares fit to the simulated curves. The fit parameters may include $\theta$, $a$, $b$, $z$, the beam waist $w_0$ and the effective coherence parameters $\mu$ or $\sigma_c$. The goodness of fit can be quantified by the root-mean-square error (RMSE) between measured and simulated normalized intensities, possibly weighted by the measurement uncertainty at each pixel. Agreement within a few percent over the central lobe and within $\sim 10\%$ over the first sidelobes would confirm the validity of the model for practical purposes.

Additionally, the partial coherence effects can be probed by introducing a rotating diffuser in one or more arms, or by deliberately detuning the optical paths beyond the coherence length to reduce the mutual coherence. By systematically varying the diffuser speed or the path-length difference, one can map the relationship between coherence parameters and fringe visibility, thereby testing the predictions of the cross-spectral-density model.

\section{Applications}
\label{sec:applications}
The three-beam double-slit configuration with edge-zero illumination leads naturally to a variety of applications in optics and photonics, thanks to its ability to engineer both the amplitude and phase across the aperture and its angular spectrum.

\subsection{Aperture apodization and spectral leakage control}
By combining several plane waves with appropriately chosen incidence angles and relative phases, the illumination across the transmitting regions can approximate classical apodization windows, such as Hann, Hamming or Blackman profiles. These windows are widely used in signal processing to reduce spectral leakage at the cost of a moderate broadening of the main lobe. In the spatial domain, this translates into reduced diffraction sidelobes, improved dynamic range for detecting weak features and cleaner separation of carrier bands in Fourier-based demodulation schemes. The edge-zero condition itself already behaves as a first-order apodizer by enforcing field nulls at the slit edges; the addition of higher-order cosine terms refines the sidelobe suppression.

\subsection{Pseudo-invariant beams and transport channels}
When an integer number of bright interference fringes is fitted inside each slit so that the field is truncated precisely at its natural zeros, the transmitted profile can behave approximately as an eigenmode of Fresnel propagation over a finite distance range. The resulting patterns preserve their shape (up to a global phase and slow scaling) upon propagation, acting as pseudo-invariant channels in one dimension. Such channels are attractive for free-space optical interconnects, alignment beams and slope-sensing arrangements where robustness to moderate longitudinal misalignment is required. Extending the concept to two-dimensional rectangular or annular pupils yields lattice-like intensity patterns that exhibit Talbot-like self-imaging within a slowly varying envelope.

\subsection{Structured illumination and super-resolution}
Structured illumination microscopy and related techniques rely on sinusoidal or multi-tone illumination patterns to shift high spatial frequencies of the object into the passband of the imaging system. The three-beam configuration considered here provides a natural generator of such carriers, with the spatial frequency $f_c = \sin\theta/\lambda$ tunable via the incidence angle. By using edge-conditioned and apodized illumination, the contrast of the structured pattern can be enhanced while minimizing spectral cross-talk and leakage, which are critical for accurate reconstruction of high-frequency components.

\subsection{Interference lithography and photonic lattices}
Multi-beam interference is a standard tool for creating periodic and quasi-periodic refractive-index patterns in photoresists and photopolymers. The double-slit envelope offers an additional degree of spatial gating, confining the exposure to selected regions and mitigating stitching artifacts near the boundaries. By extending the three-beam design to more components, it is possible to synthesize multi-tone lattices, introduce controlled defects or phase dislocations and implement graded or quasi-periodic structures for photonic devices.

\subsection{Coherent multiplexing for imaging and sensing}
Distinct incidence angles correspond to distinct carriers in the spatial-frequency domain. This enables frequency-division multiplexing in coherent imaging techniques such as digital holographic microscopy and coherent diffraction imaging. Several probe channels can be encoded and transmitted through the same pupil, then separated numerically in Fourier space. The partial coherence modeling discussed earlier provides additional knobs to trade off interference visibility against speckle noise, which is often desirable in metrology and profilometry.

\section{Conclusions}
\label{sec:conclusions}
We have presented a unified analytical and numerical framework for double-slit diffraction under three-plane-wave coherent illumination, extended to $N$ components and partially coherent sources. The edge-zero strategy yields compact closed-form Fresnel expressions, significantly simplifying the analysis while enabling intuitive spectral engineering. Numerical simulations confirm the validity of the derived expressions and highlight the flexibility of multi-wave illumination for tailoring diffraction patterns.

The methodology offers direct applicability to apodization, structured illumination, multiplexed metrology and propagation-invariant beam shaping. Future work may include experimental validation, optimization for inverse design and extension to two-dimensional apertures.

\section*{Acknowledgments}
The authors acknowledge support from their home institution and fruitful discussions with colleagues in the optics group.

\section*{Declarations of interest}
The authors declare that they have no known competing financial interests or personal relationships that could have appeared to influence the work reported in this paper.

\end{document}